\documentclass[journal=jacsat,manuscript=article]{achemso}

\usepackage[version=3]{mhchem} 
\usepackage{graphicx}

\author{Christopher J. Flower}
\email{cflower@umd.edu}
\affiliation[J]
{Joint Quantum Institute, NIST/University of Maryland, College Park, Maryland 20742, USA}
\alsoaffiliation[QTC]
{Quantum Technology Center, NIST/University of Maryland, College Park, Maryland 20742, USA}
\author{Sabyasachi Barik}
\affiliation[J]
{Joint Quantum Institute, NIST/University of Maryland, College Park, Maryland 20742, USA}
\author{Mahmoud Jalali Mehrabad}
\affiliation[J]
{Joint Quantum Institute, NIST/University of Maryland, College Park, Maryland 20742, USA}
\alsoaffiliation[QTC]
{Quantum Technology Center, NIST/University of Maryland, College Park, Maryland 20742, USA}
\author{Nicholas J Martin}
\affiliation[Sheffield]
{Department of Physics and Astronomy, University of Sheffield, Sheffield S3 7RH, UK}
\author{Sunil Mittal}
\affiliation[NE]
{Department of Electrical and Computer Engineering, Northeastern University, Boston, Massachusetts 02115, USA}
\author{Mohammad Hafezi}
\affiliation[J]
{Joint Quantum Institute, NIST/University of Maryland, College Park, Maryland 20742, USA}
\alsoaffiliation[QTC]
{Quantum Technology Center, NIST/University of Maryland, College Park, Maryland 20742, USA}
\title[]
  {Topological Edge Mode Tapering}

\abbreviations{IR,NMR,UV}
\keywords{American Chemical Society, \LaTeX}

\begin{document}

\begin{tocentry}
\includegraphics[width=8.5cm]{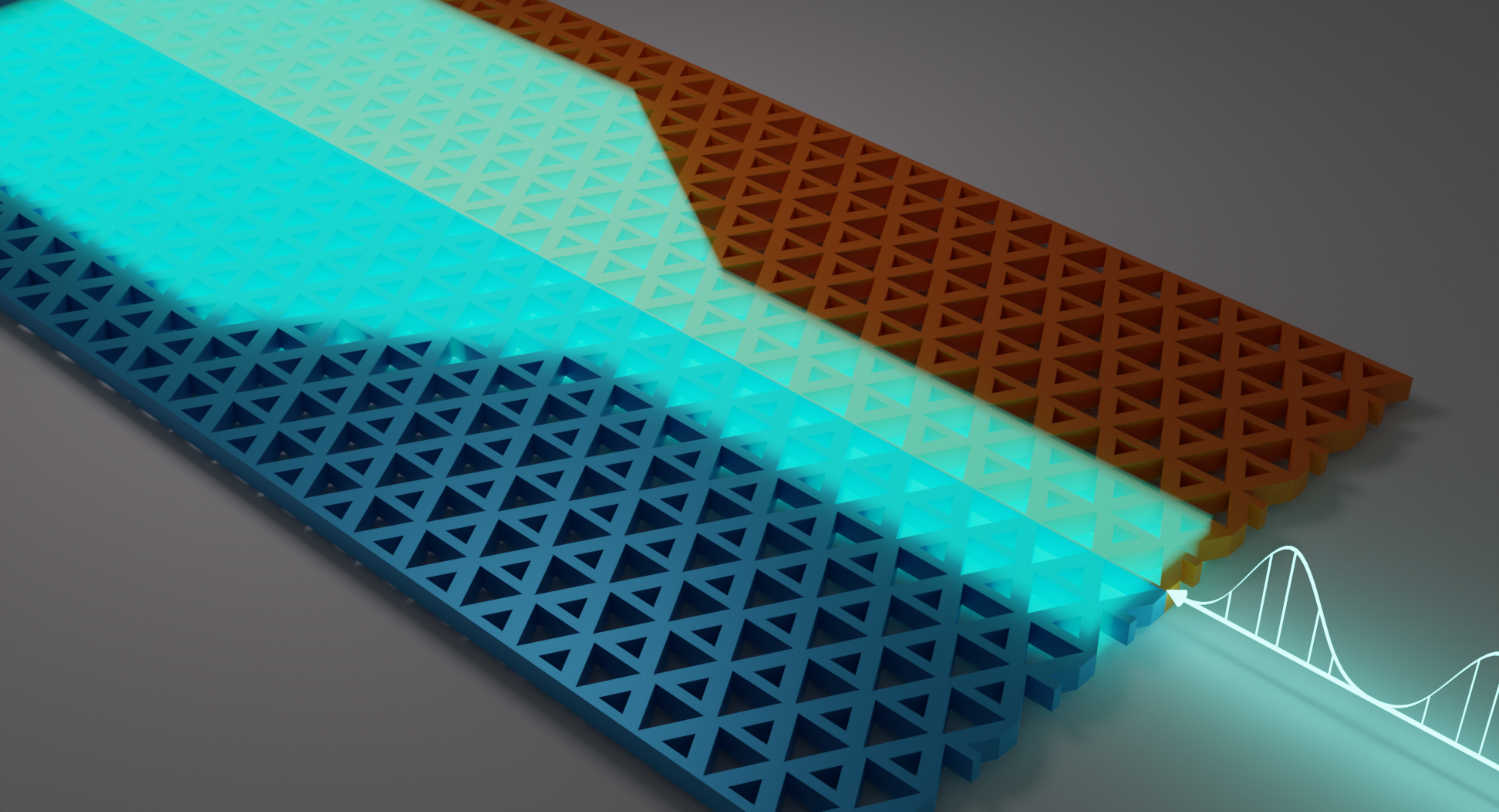}
 \label{toc}
\end{tocentry}

\begin{abstract}
Mode tapering, or the gradual manipulation of the size of some mode, is a requirement for any system that aims to efficiently interface two or more subsystems of different mode sizes. While high efficiency tapers have been demonstrated, they often come at the cost of a large device footprint or challenging fabrication due to backscattering or excitation of higher-order modes. Topological photonics, offering robustness to certain types of disorder as well as chirality, has proved to be a well-suited design principle for numerous applications in recent years. Here we present a new kind of mode taper realized through topological bandgap engineering. We numerically demonstrate a sixfold change in mode width over an extremely compact 8$\mu m$ distance with near unity efficiency in the optical domain. With suppressed backscattering and no excitation of higher-order modes, such a taper could enable new progress in the development of scalable, multi-component systems in classical and quantum optics.
\end{abstract}

\section{Introduction}
Ever since the discovery that topological physics was not limited to condensed matter systems, topology has emerged as a new design principle in optics \cite{OzawaRMP, Lu2014, Khanikaev2017, Hafezi2011, Hafezi2013, Rechtsman2013}. Promising built-in protection against defects and disorder, as well as chirality, topology in optics has led to the development of novel ideas including topological delay lines \cite{Hafezi2011}, lasers \cite{StJean2017,Bahari2017, Bandres2018, Zhao2018, Ota2018, Yang2022}, waveguides \cite{Barik2018}, antennas \cite{gorlach2018far}, fibers \cite{Lu2018}, resonators \cite{Barik2020,Mehrabad2020,Mehrabad2023,MJM_Opt_2020}, sources of quantum light \cite{Mittal2018}, devices enabling robust routing of photons \cite{BlancoRedondo2018,Zhao2019}, and more.

At the same time, mode matching is a ubiquitous problem in the field of optics, and often requires the manipulation of the spatial extent of some mode. Mode engineering of this kind is often employed in optical systems to increase coupling efficiency between a large optical source and some sample \cite{Almeida2003, Mu20, Dutta16}. It is notably a crucial component for the development of hybrid photonic integrated systems, which offer a path to solve several outstanding engineering challenges in both classical and quantum photonics \cite{Heck2013, Kim2017, Elshaari2020}. Additionally, it is a principle often employed in the development of hybrid quantum systems, where coupling efficiency between component systems can be limited due to impedance mismatch \cite{Kurizki15}. For example, a taper may be used to couple a many-wavelength scale transducer to a waveguide with a much smaller cross-section \cite{Dahmani2019}. In addition to improving coupling, a wider or more distributed mode may be advantageous for applications that require careful power management. While mode tapering has been extensively studied, modern designs must carefully avoid backscattering or excitation of higher order modes. It therefore remains desirable to design new, compact, and efficient mode tapers without these limitations \cite{Almeida2003, Mu20, Dutta16}.

 \begin{figure}[htbp]
\centering\includegraphics[width=3.25in]{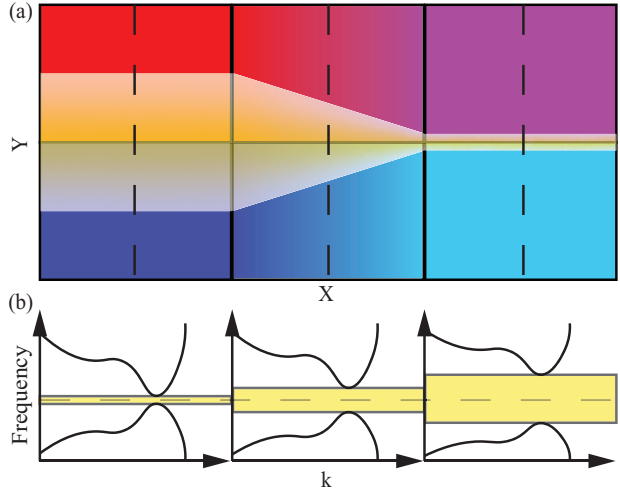}
 \caption{The relationship between bandgap width and edge mode width. a) A representation of the interface of two topologically distinct materials. An edge mode exists at the interface of the top and bottom regions, but the width of that mode is determined by a changing control parameter, represented by the color gradient which changes from left to right. b) The corresponding bulk band structures at different points in space.} 
 \label{Schematic}
\end{figure}

 \begin{figure}[htbp]
 \centering\includegraphics[width=7in]{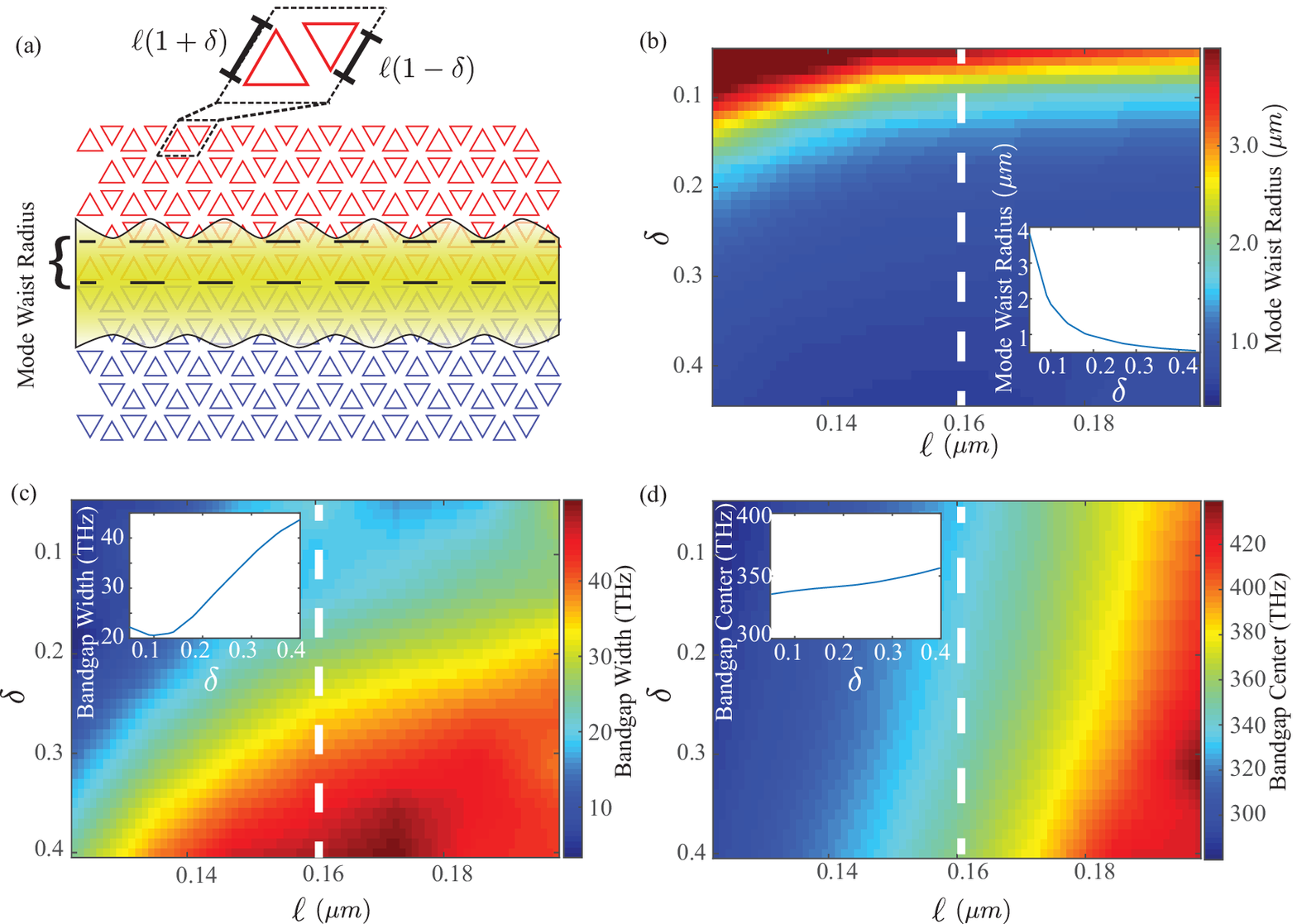}
 \caption{\label{F2} The relationship between lattice parameters, $\ell$ and $\delta$, and properties of the topological photonic crystal waveguide system. a) Schematic representation of the valley-Hall topological photonic crystal design and edge mode. Lattice parameters are indicated.  b-d) Mode waist radius, bulk bandgap width, and center frequency vs lattice parameters $\ell$ and $\delta$. Each panel includes an inset displaying a cross section of the data at $\ell=$0.16 $\mu m$. All data is interpolated for clarity.}
\end{figure}

In this work, we describe a novel means of mode tapering through topological bandgap engineering. By taking advantage of the robust,  unidirectional, and fundamentally single-mode nature of topological edge modes, we present a taper that allows abrupt many-fold enhancement or reduction of mode size with negligible losses. We achieve this by taking advantage of the inverse relationship between the topological bandgap and edge mode width, schematically shown in Fig.~\ref{Schematic}. By spatially manipulating the lattice parameters (introduced in Fig.~\ref{F2}a) we are able to tune the bulk bandgap of a valley-Hall topological photonic crystal waveguide \cite{Ma2016, Dong2017, Chen2017, He2019, Noh2018, Shalaev2019, Barik2020,Mehrabad2023,MJM_Opt_2020}, and by extension the edge mode width, while maintaining a constant center frequency. While this approach can be applied to any frequency in the electromagnetic spectrum, we focus in the optical domain for concreteness and demonstrate a bandwidth of 9 THz around a central frequency of 353 THz. Using 2D finite difference time domain techniques, we demonstrate an ultra-efficient ($>$99.5\%) 6-fold mode taper over an ultra-short distance ($<$8 $\mu m$ near a wavelength of $850 nm$) without the use of any optimization techniques. Furthermore, we stress that the principle of this approach to robust mode tapering is broadly applicable and could be realized in other platforms, such as acoustic, electronic, or atomic systems, where topological physics is studied.

\section{Topological Bandgap Engineering for Tapering}
Fig. \ref{Schematic} illustrates the expected relationship between edge mode confinement and topological bandgap, as well as how it may be leveraged to engineer a mode taper. In particular, we expect the confinement length of the topological edge states, which are exponentially localized at the boundary, to scale roughly inversely with the bulk bandgap. Panel a depicts an edge mode at the interface of two topologically distinct regions (top and bottom, here indicated by hot and cold color schemes online). A left to right gradient in both the top and bottom regions represents the changing of some control parameter that determines the width of the bulk topological bandgaps. We stipulate that the control parameters in the system are such that the bulk bandgaps coincide in frequency in the top and bottom regions at the same point along the interface. These bulk bandgaps are illustrated in panel b at three points along the taper, indicated by dotted lines in panel a. 

\section{Photonic Crystal Design}

The topological photonic crystal system that is the subject of this work is based on a valley-Hall design composed of a honeycomb lattice of triangular holes \cite{Barik2020, Dong2017, Chen2017, He2019, Noh2018, Shalaev2019, Ma2016,Mehrabad2023,MJM_Opt_2020}. The rhombic unit cell of the crystal in question consists of two triangular holes of side length $\ell(1 \pm \delta)$, shown in Fig. \ref{F2}a. The lattice constant is chosen to be 0.46 $\mu m$. Time reversal symmetry mandates that the Berry curvature integrated over the full Brillouin zone is zero, but crucially the Berry curvatures at the $K$ and $K'$ points are of opposite sign. Interfacing such a region with its mirror image results in a pair of helical modes at the boundary.

These edge modes can be characterized by a physical width, or mode waist radius, a bandgap width, and a bandgap center frequency. The mode waist radius is defined here as the distance over which the field amplitude decays from its maximum to $1/e$ times its maximum in the in-plane direction orthogonal to the propagation direction of the mode. If both regions have the same lattice parameters ($\ell$ and $\delta$) the mode waist radius will be the same on both sides of the boundary. The bandgap width and center frequency are bulk properties of the two topologically distinct regions that determine the frequency and bandwidth of the edge mode.

\section{Effective Hamiltonian and Analysis}

To get an approximate idea of how these properties depend on the lattice parameters $\ell$ and $\delta$, we can turn to the Hamiltonian describing the valley-Hall topological photonic crystal insulator near each of the $K$ and $K'$ valleys \cite{Ma2016}, given by:

\begin{equation}\label{eq:1}
H_{\rm eff} = v_D(\tau_z \sigma_x \delta k_x+\tau_0\sigma_y \delta k_y)+ m\tau_0\sigma_z
\end{equation}
where $v_D$ is the group velocity and $\delta \mathbf{k}$ are momenta relative to the $K/K'$ points. The Pauli matrices $\sigma_{x,y,z}$ and $\tau_{x,y,z}$ act on the basis of right and left circularly polarized eigenmodes of the unperturbed lattice and the valley degree of freedom, respectively. The perturbation or mass term, $m$, is calculated by determining the overlap of the change in permittivity of the perturbed crystal with the field intensities of the eigenmodes of the unperturbed crystal.

When one considers an interface at $y=0$ with opposite masses $\pm m$ on either side, Eq.1 yields a solution state that is bound at $y=0$, and freely propagates along the x-axis. Apart from the four-component spinor part, the spatial part takes the form:

\begin{equation}\label{eq:2}
\Psi(x,y) \propto e^{-\frac{m|y|}{v_D}} e^{\pm \frac{ik_x x}{v_D}}.
\end{equation}
As can be seen, the decay length away from the interface is proportional to the inverse of the mass term, and by extension the bandgap. 
In particular, the change in permittivity that determines the value of $m$ is governed by the change in the area of each triangle in the unit cell. This quantity goes to zero as $\delta$ goes to zero, but also depends on $\ell$.

\section{Simulation of Bulk and Edge Mode Properties} In order to create a functional tapered waveguide, we must carefully choose the lattice parameters such that as we move along the interface the bandgap width and the mode waist radius change while the center frequency remains the same. The relationships between each of these mode properties and the lattice parameters, $\ell$ and $\delta$, are displayed in Fig. \ref{F2}b-d, where each respective inset shows a cross section of the data with $\ell$ fixed at 0.16 $\mu m$. As hypothesized, the dependency of the mode waist radius on the lattice parameters is roughly inverse that of the bandgap width. The edge mode becomes more and more tightly confined and the bandgap grows as the difference between the triangular holes in a single rhombic unit cell is increased. The bandgap center frequency on the other hand depends only very weakly on $\delta$, in agreement with prediction from tight-binding models.

\begin{figure}[htbp]
\centering\includegraphics[width=3.25in]{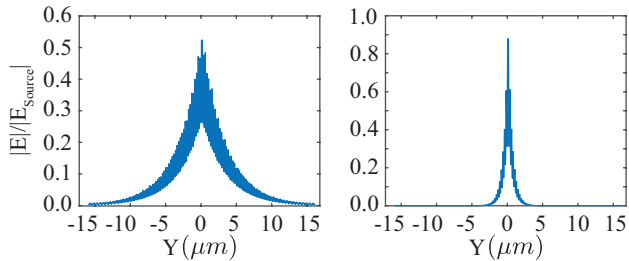}
\caption{\label{F3} E field amplitudes versus in-plane displacement (Y) across the interface of the topologically distinct regions of the valley-Hall photonic crystal. a) $\delta= 0.05$ and $\ell=$ 0.17 $\mu m$. b) $\delta= 0.40$ and $\ell=$ 0.16 $\mu m$. }
\end{figure}

With these landscapes in hand, we can now design tapered topological photonic crystal waveguides across a wide range of operational frequencies and mode waist diameters. Fig. \ref{F3} displays cross sectional snapshots of the mode profiles of two points in this parameter space. In particular, panel a shows the electric field amplitudes of the edge mode for $\ell = $ 0.17 $\mu m$ and $\delta = $ 0.05, while panel b shows that for $\ell = $ 0.16 $\mu m$ and $\delta = $ 0.40, both at their center frequency of 353 THz. The mode waist radius is determined by fitting a straightforward envelope function and gives 3.5 and 0.6 $\mu m$ respectively. 

\begin{figure}[htbp]
\centering\includegraphics[width=7in]{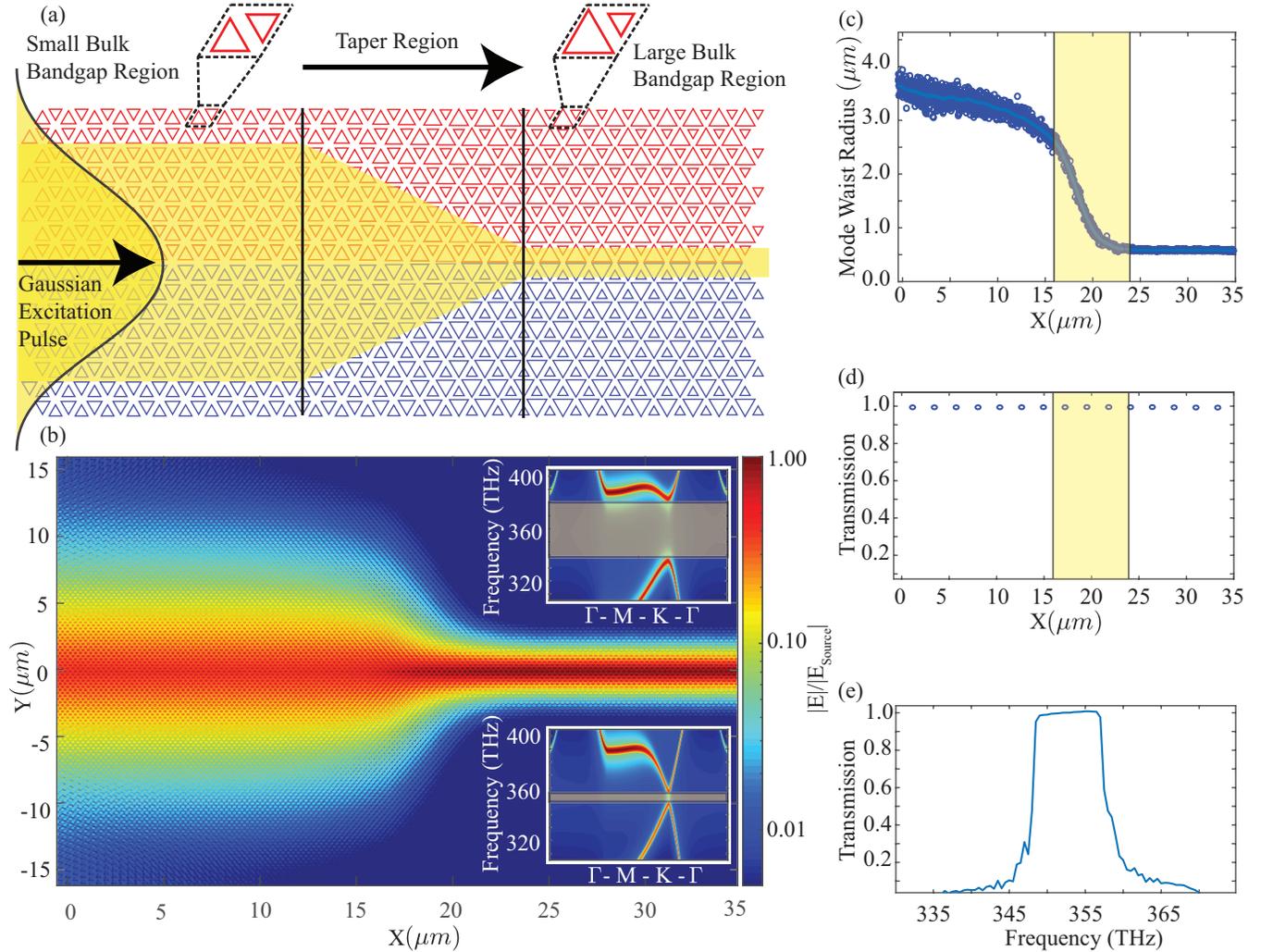}
\caption{\label{F4} The topologically tapered waveguide system. a) A schematic representation of the fully tapered topological photonic crystal system and excitation scheme. In simulation, this topological photonic crystal was excited with a Gaussian beam source of approximately 5 $\mu m$ mode waist radius. b) The magnitude of the field amplitude for the tapered system, normalized by the source magnitude $(|E|/|E_{Source}|)$, presented on a logarithmic scale. The bottom and top insets show bulk bandstructures for the lattice parameters used to the left and right of the taper region, respectively. c) Mode waist radius along the direction of propagation. The region in which the photonic crystal parameters are tapered is indicated. d) Transmission, defined as the integral of the real part of the Poynting vector over a surface orthogonal to the propagation direction of the mode, along the optical path of the waveguide. Normalization is with respect to transmission at the far left of the system, but downstream from the source. e) Transmission band of the full system. The bandwidth or full width at half maximum is approximately 9 THz centered around 353 THz.
 }
\end{figure}

\section{Simulation of Tapered Edge Mode Properties} Using these two points in parameter space as endpoints, we have designed a tapered photonic crystal consisting of a small bandgap region ($\ell = $ 0.17 $\mu m$ and $\delta = $ 0.05), a linear taper region, and a large bandgap region ($\ell = $ 0.16 $\mu m$ and $\delta = $ 0.40). In the small and large bandgap regions the lattice parameters remain fixed, while in the taper region the lattice parameters are linearly tuned between the two. The resulting design and simulation scheme is displayed in Fig. \ref{F4}a. By exciting the guided edge mode from the left with a 60 THz wide Gaussian excitation pulse we are able to determine the mode structure and properties of the tapered topological waveguide system. Fig. \ref{F4}b shows the calculated field profile of the guided mode at 353 THz, clearly showing the dramatic decrease in mode waist radius along the propagation direction. This effect is shown quantitatively in Fig. \ref{F4}c, demonstrating an approximately 6-fold reduction in mode waist radius (or alternatively a 5.8 $\mu m$ reduction in mode waist diameter) through the taper region, indicated by vertical lines. 

To ensure that such an abrupt change in the mode profile hasn't been achieved at the cost of reduced efficiency, we present transmission data through the tapered system in Fig. \ref{F4}d at 353 THz. Based on this data we can bound the mode conversion loss of the taper at $<$ 0.5$\%$ over the 8 $\mu m$ taper. This data is normalized by transmission data from an untapered topological photonic crystal waveguide matching the parameters of the small bandgap system, or the wider, left-hand side of the tapered system. 

In order to be a useful component in a wide range of applications, such as those focused on quantum or classical information processing, we must also guarantee that the bandwidth of our device is sufficiently large. The frequency response of the tapered waveguide is displayed in Fig.\ref{F4}e which shows the bandwidth to be approximately 9 THz, centered around 353 THz.

\section{Discussion} Although we have only considered linear tapers and this system is entirely unoptimized, we have nevertheless shown that this means of mode tapering in a topological system can exhibit significant mode width engineering with extremely high efficiency over an ultra-compact distance. In particular, we have demonstrated a change in mode waist diameter of 5.8 $\mu m$ over an 8 $\mu m$ taper length while suffering only $<0.5\%$ mode conversion loss. This degree of efficiency is made possible by the inherent topological robustness of the mode and the absence of any higher-order modes, regardless of mode width. For comparison, adiabatic linear tapers in silicon-on-insulator waveguides (which lack these favorable properties) may achieve comparable losses at only extremely shallow taper angles (1$^\circ$). In such a system, a few micron increase in mode waist diameter with negligible loss would require hundreds of microns of taper length \cite{Sheng12}. This device footprint can be reduced with the use of more sophisticated techniques, such as an inverse parabolic taper, but this still requires 40 $\mu m$ of taper length for a change in mode waist diameter of $\approx$ 4.5 $\mu m$, and suffers mode conversion losses of greater than 5\% \cite{Almeida2003}. 

Such topological systems are therefore promising candidates for robust and efficient interfaces between physical systems of varying mode sizes. Uses for an interface of this nature are plentiful, ranging from simple applications like increasing the coupling efficiency between optical systems \cite{Almeida2003, Mu20, Dutta16}, to more involved ones such as the development of hybrid integrated photonics for quantum information processing \cite{Heck2013, Kim2017, Elshaari2020}. Furthermore, as this approach is not fundamentally limited to photonic systems, it may even be useful in the development of more general hybrid quantum systems that may suffer from impedance mismatch due to different mode sizes \cite{Kurizki15}.

\begin{suppinfo}
Additional studies of coupling a valley-Hall topological photonic waveguide to a nanobeam waveguide and topological robustness for different mode waist radii.
\end{suppinfo}

\begin{acknowledgement}
This research was supported by The Office of Naval Research ONR-MURI grant N00014-20-1-2325, AFOSR FA95502010223, NSF PHY1820938, NSF DMR-2019444, and The Army Research Laboratory grant W911NF1920181.

The authors thank Wade  DeGottardi, Luke Wilson, and Amir Safavi-Naeini for  their helpful  discussions. The authors declare no conflicts of interest.
\end{acknowledgement}

\bibliography{TaperBib}

\end{document}